\documentstyle[aps,preprint]{revtex}

\widetext
\draft
\tighten

\begin{document}

\preprint{EFUAZ FT-95-13}

\title{Extra Dirac Equations\thanks{Submitted
to ``{\it Journal of Mathematical Physics}".}}

\author{{\bf Valeri V. Dvoeglazov}\thanks{On leave of absence from
{\it Dept. Theor. \& Nucl. Phys., Saratov State University,
Astrakhanskaya ul., 83, Saratov\, RUSSIA.}\,
Email: dvoeglazov@main1.jinr.dubna.su}}

\address {
Escuela de F\'{\i}sica, Universidad Aut\'onoma de Zacatecas \\
Antonio Doval\'{\i} Jaime\, s/n, Zacatecas 98000, ZAC., M\'exico\\
Internet address:  VALERI@CANTERA.REDUAZ.MX
}

\date{April 15, 1995}

\maketitle

\begin{abstract}
This paper has rather a pedagogical meaning.
Surprising symmetries in the $(j,0)\oplus (0,j)$
Lorentz group  representation space  are analyzed.
The aim is to draw reader's attention to the possibility
of describing the particle world on the ground of
the Dirac ``doubles". Several tune points of the variational
principle for this kind of equations are briefly discussed.
\end{abstract}


\newpage

Chiral invariance has profound significance in the modern
theory of weak interactions and it has played a prominent
role in understanding the low
energy properties of strong interactions. But,
it has long been recognized that the Dirac Lagrangian
for massive fermions is invariant only under simultaneous chiral
transformation $\psi\rightarrow \gamma_5 \,\psi$
and ``mass reversal" transformation
$m\rightarrow -\,m$. The idea of construction of particle dynamics
on the ground of two Dirac equations (the second one has
the opposite sign at the mass term) has been proposed shortly after
an appearance of the famous
equation~\cite{Markov,Beli}.\footnote{More exactly, in the papers of 1937
Prof. M. A. Markov considered the second-order equation
for 4-component spinor. But, in the latest work  he has shown
that it is equivalent to the first order equation for
the eight-component  function (or to the set of two Dirac
equations).} Unfortunately, both those investigations and
the papers~\cite{Brana} have been forgotten.
Another possibility of two extra Dirac equations (two extra
``square roots" of the Klein-Gordon equation) for
4-spinors of the second kind~\cite{Cartan} seems also to have
escaped from  attention of theoreticians.  The only mention of this
possibility, that I was able to find, is in ref.~\cite[Eq.(8)]{Sokolik},
in connection with an ``anomalous" representation of the inversion
group~\cite{Gelfand}. Some speculations on the possible relevance
of such the kind of equations to description of neutrino and
on the eventual connection with existence  of isotopic spin
have been presented there. In this essay I am going to undertake
a detailed analysis of the Dirac ``doubles"\,\footnote{The meaning
of this terminology  will become clear in the sequel.} and to found
some relations with the models discussed
recently~\cite{DVA,DVO,DVAG0,DVAG}.

Beforehand, I would like to reproduce here
the way of deriving the usual Dirac equation on the ground of
the Wigner rules~\cite{Wigner} of Lorentz transformations
of the $(0,j)$ left $\phi_L (p^\mu)$ and  the $(j,0)$ right
$\phi_R (p^\mu)$ spinors:
\begin{mathletters}
\begin{eqnarray}
(j,0):&&\quad\phi_R (p^\mu)\, = \,\Lambda_R (p^\mu \leftarrow
\overcirc{p}^\mu)\,\phi_R (\overcirc{p}^\mu) \, = \, \exp (+\,{\bf J} \cdot
{\bbox \varphi}) \,\phi_R (\overcirc{p}^\mu)\quad,\\
(0,j):&&\quad \phi_L (p^\mu)\, =\, \Lambda_L (p^\mu \leftarrow
\overcirc{p}^\mu)\,\phi_L
(\overcirc{p}^\mu) \, = \, \exp (-\,{\bf J} \cdot {\bbox \varphi})\,\phi_L
(\overcirc{p}^\mu)\quad.\label{boost0}
\end{eqnarray}
\end{mathletters}
It is known and is given, {\it e.g.}, in the
papers~\cite[p.9]{Faustov},~\cite[p.43-44]{Ryder}.  In
ref.~[10b,footnote \# 1] several important points have been shown at:
``Refer to Eqs. (1a) and (1b) and set ${\bf J} = {\bbox \sigma}/2$...
Spinors [implied by the arguments based on parity symmetry and that
Lorentz group is essentially $SU_R (2) \otimes SU_L (2)$] turn out to be
of crucial significance in constructing a field $\Psi (x)$ that describes
eigenstates of the Charge operator, $Q$, if [in the rest]
\begin{equation}\label{rb}
\phi_R (\overcirc{p}^\mu)
=\pm \phi_L (\overcirc{p}^\mu)
\end{equation}
(otherwise physical eigenstates are no longer
charge eigenstates). We call [this relation], the ``Ryder-Burgard
relation"... Next couple the Ryder-Burgard relation with Eqs.
(1a) and (1b) to obtain
\begin{eqnarray}
\pmatrix{\mp m \,\openone & p_0 + {\bbox \sigma}\cdot {\bf p}\cr
p_0 - {\bbox \sigma}\cdot {\bf p} & \mp m \,\openone\cr} \psi (p^\mu)\,
=\, 0\quad.
\end{eqnarray}
[Above we have used the property $\left [\Lambda_{L,R} (p^\mu
\leftarrow \overcirc{p}^\mu)\right ]^{-1} =
\left [\Lambda_{R,L} (p^\mu \leftarrow \overcirc{p}^\mu)\right ]^\dagger$
and that both ${\bf J}$ and $\Lambda_{R,L}$ are Hermitian
for the finite $(j=1/2,0)\oplus (0,j=1/2)$ representation
of the Lorentz group.]
Introducing $\psi (x) \equiv \psi (p^\mu)  \exp (\mp ip\cdot x)$
and letting $p_\mu \rightarrow i\partial_\mu$, the above equation
becomes: $(i\gamma^\mu \partial_\mu - m \,\openone) \psi (x) = 0$.
This is the Dirac equation for spin-1/2 particles with
$\gamma^\mu$ in the Weil/Chiral representation."
In the standard (generalized canonical)
representation such a choice of the Faustov-Ryder-Burgard
relation\footnote{More general form of the relation (\ref{rb})
has been given in the unpublished preprint of
Prof. Faustov~\cite[Eq.(22a)]{Faustov}.}
and the use of a representation of the ${\bf J}$ matrices in which $J_z$
is diagonal imply the well-known spinorial basis of bispinors $u_h$ and
$v_h$ in the $(j,0)\oplus (0,j)$ representation space\cite{DVA}:
\begin{eqnarray}
u_{+j} (\overcirc{p}^\mu) =\pmatrix{N(j)\cr 0\cr . \cr .
\cr . \cr 0\cr}\quad,\quad
u_{j-1} (\overcirc{p}^\mu)=\pmatrix{0\cr
N(j)\cr . \cr .  \cr . \cr 0\cr}\quad,\quad \ldots,\quad
v_{-j} (\overcirc{p}^\mu)=\pmatrix{0\cr 0\cr . \cr . \cr . \cr N(j)\cr}\quad.
\end{eqnarray}
The normalization factor is convenient to choose $N(j)=m^j$ in order
the rest spinors to vanish in the massless limit.

The pioneer study of the $(j,0)\oplus (0,j)$ representation space
has been undertaken by Weinberg~\cite{Weinberg}
in the sixties\footnote{Of course, the $j=1/2$ Dirac fermions are also
contained in this scheme.}.
The use of the Faustov-Ryder-Burgard relation (\ref{rb})
in the case of $j=1$ permits
us to obtain the Weinberg equation; exactly, its modified
version~\cite{DVA}:
\begin{equation}\label{mweq}
\left (\gamma^{\mu\nu}\partial_\mu \partial_\nu + \wp_{u,v} m^2 \right )
\psi (x) = 0
\end{equation}
with $\wp_{u,v} = \pm 1$ and $\gamma_{\mu\nu}$ are the
Barut-Muzinich-Williams $j=1$ matrices~\cite{Barut}.
A boson described by Eq. (\ref{mweq}) has the opposite relative
intrinsic parity with respect to its antiboson. This is
an example of the Bargmann-Wightman-Wigner (BWW) type
quantum field theory~\cite{Wigner2}.\footnote{About connections
of this type of Poincar\'e invariant theories with
the constructs proposed by Foldy and Nigam~\cite{Nigam} and
by Gelfand, Tsetlin and Sokolik~\cite{Gelfand,Sokolik}
even before an appearance ref.~\cite{Wigner2} see
ref.~\cite{DVAG,DVO951,DVO952}.} An explicit construct of
this case of the BWW theories has been proposed recently~\cite{DVA}
by Ahluwalia (see also ref.~\cite{Sila}) and has been analyzed
in ref.~\cite{DVO}.

Now, let me utilize the Faustov-Ryder-Burgard relation in a slightly
different form. Namely, let assume that right-  and left-
complex-valued spinors are connected in the rest frame
in the following way:
\begin{equation}\label{rbn}
\phi_R (\overcirc{p}^\mu) = \pm i\,\phi_L (\overcirc{p}^\mu)\quad.
\end{equation}
In fact, this choice (\ref{rbn})  corresponds to the following
spinorial basis of
bispinors\footnote{Here and below I work in
the chiral representation of $\gamma$ matrices.}
(provided that 2-spinors are chosen as $\phi_{R}^{+}
(\overcirc{p}^\mu) = column (1\quad 0)$ and
$\phi_{R}^{-} (\overcirc{p}^\mu) = column (0 \quad 1)$):
\begin{mathletters}
\begin{eqnarray}
\Upsilon_+ (\overcirc{p}^\mu) \,&=&\,
\sqrt{{m\over 2}}\pmatrix{1\cr 0\cr -i\cr
0\cr}\quad,\quad
\Upsilon_- (\overcirc{p}^\mu) = \sqrt{{m\over 2}}
\pmatrix{0\cr 1\cr 0\cr -i\cr}\quad,\quad\\
{\cal B}_+ (\overcirc{p}^\mu) \,&=&\,
\sqrt{{m\over 2}}\pmatrix{1\cr 0\cr i\cr 0\cr}\quad,\quad
{\cal B}_- (\overcirc{p}^\mu) =
\sqrt{{m\over 2}}\pmatrix{0\cr 1\cr 0\cr i\cr}\quad.
\end{eqnarray}
\end{mathletters}
If couple Eq. (\ref{rbn}) with (1a) and (1b) one obtains
the equations in the momentum representation:\footnote{If
accept the way of deriving the equations satisfied
by $(j,0)\oplus (0,j)$ spinors on the base of
the Wigner rules (1a) and (1b) in the instant-form of field theory
we should assume that  $m\neq 0$.
Therefore, the framework of the front-form  relativistic
dynamics~\cite{Dirac0,DVALF,DVAG0} could be more convenient
for study of the massless limit. We are going to regard
this question in following publications.}
\begin{mathletters}
\begin{eqnarray}
\left [i\gamma^5 \hat p - m\right ]\Upsilon_{+,-} (p^\mu) = 0\quad,
\label{eq1}\\
\left [i\gamma^5 \hat p + m\right ]{\cal B}_{+,-} (p^\mu) =0\quad,
\label{eq2}
\end{eqnarray}
\end{mathletters}
satisfied by the 4-spinors of the second kind:
\begin{mathletters}
\begin{eqnarray}
\Upsilon_+ (p^\mu) &=& {1\over 2\sqrt{p_0+m}}\pmatrix{p^{+}+ m \cr p_r\cr
-i\,(p^- + m)\cr  i\,p_r\cr} = \gamma^5 {\cal B}_+ (p^\mu)
= \gamma^5 S^c_{[1/2]} \, \Upsilon_- (p^\mu)\quad,\quad\label{s1}\\
\Upsilon_- (p^\mu) &=& {1\over 2\sqrt{p_0+m}}
\pmatrix{p_l\cr p^- +m\cr i\,p_l\cr  -i\,(p^+ + m)\cr}=
\gamma^5 {\cal B}_- (p^\mu)
= -\gamma^5 S^c_{[1/2]}\,  \Upsilon_+ (p^\mu)\quad,\quad\label{s2}\\
{\cal B}_+ (p^\mu) &=& {1\over 2\sqrt{p_0+m}}
\pmatrix{p^+ +m\cr p_r\cr i\,(p^- +m)\cr  -i\,p_r\cr}=
\gamma^5 \Upsilon_+ (p^\mu) = -\gamma^5 S^c_{[1/2]}\,  {\cal B}_- (p^\mu)
\quad,\quad\label{s3}\\
{\cal B}_- (p^\mu) &=& {1\over 2\sqrt{p_0+m}}
\pmatrix{p_l\cr p^- +m\cr -i\,p_l\cr  i\,(p^+ + m)\cr}=
\gamma^5 \Upsilon_- (p^\mu) =\gamma^5 S^c_{[1/2]}\,
{\cal B}_+ (p^\mu)\quad.\label{s4}
\end{eqnarray}
\end{mathletters}
We have used above the following notation:
$p_r =p_x +ip_y$, $p_l =p_x -ip_y$, $p^\pm = p_0 \pm p_z$ and
\begin{eqnarray}
S^c_{[1/2]} = \pmatrix{0 & i\Theta_{[1/2]}\cr
-i\Theta_{[1/2]} & 0\cr}{\cal K}\quad,
\end{eqnarray}
with ${\cal K}$ being the operation of complex conjugation; and
$\left (\Theta_{[j]}\right )_{h,\,h^\prime} = (-1)^{j+h}
\delta_{h^\prime,\,-h}$  being the Wigner time-reversal
operator. Using the properties of 4-spinors with respect to
chiral $\gamma^5$ transformations the equations (\ref{eq1},\ref{eq2})
in the momentum representation could also be written:
\begin{mathletters}
\begin{eqnarray}
i\hat p \,\Upsilon (p^\mu) - m \,{\cal B} (p^\mu) &=& 0\quad,\quad\\
i\hat p \,{\cal B} (p^\mu) + m \,\Upsilon (p^\mu) &=& 0\quad.
\end{eqnarray}
\end{mathletters}
The bispinors (\ref{s1}-\ref{s4}) obey the normalization conditions:
\begin{mathletters}
\begin{eqnarray}
\overline \Upsilon_h (p^\mu) {\cal B}_{h^\prime} (p^\mu) \,&=&
\,- \,\overline {\cal B}_h (p^\mu) \Upsilon_{h^\prime} (p^\mu)
\,= \,im \,\delta_{h\,h^\prime}\quad,\\
\overline \Upsilon_h (p^\mu) \Upsilon_{h^\prime} (p^\mu) \,&=&\,
\overline {\cal B}_h (p^\mu) {\cal B}_{h^\prime} (p^\mu) \,=\, 0\quad,\\
\Upsilon_h^\dagger (p^\mu) \Upsilon_{h^\prime} (p^\mu)\, &=&\,
{\cal B}_h^\dagger (p^\mu) {\cal B}_{h^\prime} (p^\mu)
\,=\, p_0 \,\delta_{h\,h^\prime}\quad,\\
\Upsilon_h^\dagger (p^\mu) {\cal B}_{h^\prime} (p^\mu)\, &=&\,
{\cal B}_h^\dagger (p^\mu) \Upsilon_{h^\prime} (p^\mu) \,=\, \cases{
+ p_3  \quad,\quad\mbox{if} \quad h= +, h^\prime=  + & $ $\cr
- p_3  \quad,\quad\mbox{if} \quad h= -, h^\prime= -  & $ $\cr
p_l  \quad,\quad \mbox{if} \quad h= + , h^\prime= -  & $ $\cr
p_r  \quad,\quad \mbox{if} \quad h= - , h^\prime= +  & $ $}
\end{eqnarray}
\end{mathletters}
The properties of 4-spinors of the second kind under space inversion
are different from the 4-spinors of the first
kind:\footnote{Of course, the reader is right to ask the question:
what could we obtain if define the parity
operator according to the anomalous representation of the inversion
group, ref.~\cite{Gelfand,Sokolik}? Within the framework of this essay
we are still going to accept the viewpoint of Prof.
Ahluwalia, ref.~\cite{DVAG}: the operator of parity,
charge conjugation and time reversal
do not depend on a specific wave equation, $S^s_{[1/2]} =
e^{i\theta^s_{[1/2]}}\,\gamma_0$. ``Without this being true we would not
even know how to define charge self/anti-self conjugate $(j,0)\oplus (0,j)$
spinors."}
\begin{mathletters}
\begin{eqnarray}
S^s_{[1/2]} \Upsilon_h (p^{\prime\,\mu})\, &=&\, - i\, {\cal B}_h
(p^\mu)\quad, \label{pc1}\\
S^s_{[1/2]} {\cal B}_h (p^{\prime\,\mu}) \,&=&\, + i \,
\Upsilon_h (p^\mu)\quad.\label{pc2}
\end{eqnarray}
\end{mathletters}
In the coordinate representation the obtained equations
(\ref{eq1},\ref{eq2}) yield\footnote{
Taking into account the unusual properties of the 4-spinors
$\Upsilon_\pm$ and ${\cal B}_\pm$ with respect to the parity conjugation
(\ref{pc1},\ref{pc2}) we conclude that at the classical level
Eq. (\ref{eq}) can be put
in the form ($x^\prime = (x_0, - {\bf x})$, $\wp_{u,v}=\pm 1$)
\begin{equation}
i\hat \partial_x \Psi (x) - \wp_{u,v} m\gamma_0 \Psi (x^\prime) = 0\quad,
\end{equation}
or
\begin{equation}
i\hat \partial_{x^\prime} \Psi (x^\prime) - \wp_{u,v} m \gamma_0
\Psi (x) =0\quad.
\end{equation}
Investigations of physical consequences following from these
equations we leave for future publications. An interesting paper
in this direction is ref.~\cite{Bruce}.}
\begin{equation}
\left [\gamma^5 \gamma^\mu \partial_\mu +m \right ]\Psi (x) = 0\quad,
\label{eq}
\end{equation}
provided that $\Upsilon_h (p^\mu)$ are regarded as positive-energy solutions
and ${\cal B}_h (p^\mu)$, as negative-energy solutions.\footnote{For
the moment let us still note that we do not have  a strong theoretical
principle for the used interpretation of the 4-spinors. The determinants
of both Eq. (\ref{eq1}) and Eq. (\ref{eq2}) (from which we find
the dispersion relations) provide two signs of the energy $E=\pm
\sqrt{{\bf p}^2 +m^2}$. If assume that $\Upsilon_h (p^\mu)$ (and
${\cal B}_h (p^\mu)$) correspond to the
negative (positive) energies we have to use the equation with
the opposite sign at the mass term in the coordinate
representation. The same situation exists in the usual Dirac equation,
refs.~\cite{Markov,Beli,Sokolik,Brana}. {\it E. g.}, Prof. M. Markov
intended to utilize these two types of the Dirac fields for explanation of
mass difference between muon and electron. With an appearance of
indications at the third family of leptons (quarks) this idea has been
forgotten.}
By means of simple calculations one can derive the adjoint equation:
\begin{equation}
\overline \Psi (x) \left [ \gamma^5 \gamma^\mu \loarrow{\partial}_\mu +m
\right ] = 0\quad.\label{eqa}
\end{equation}
The equation (\ref{eq}) has been
discussed in the old literature~\cite{Sokolik} and, recently, has been
derived~\cite{DVO951,DVO952}
from the consideration of the Majorana-McLennan-Case
construct~\cite{Majorana,Case}
with self/anti-self charge conjugate spinors, developed
by Ahluwalia~\cite{DVAG0,DVAG}.
The equation is connected by unitary transformation
${\bf U} = (1 - i\gamma_5)/\sqrt{2}$ with the usual
Dirac equation:\footnote{
For the sake of completeness let us note that a Pauli term,
that could describe interactions of the particle possessing an
anomalous magnetic moment, is invariant with respect to the
transformation: ${\bf U}^{-1} \sigma^{\mu\nu} {\bf U} = \sigma^{\mu\nu}$.}
\begin{eqnarray}
\lefteqn{\left [i \gamma^\mu \partial_\mu  - m \right ]
\psi (x) = 0 \quad \rightarrow \quad}\label{eqd}\\
&\rightarrow&\quad\left [i {\bf U}^{-1}
\gamma^\mu {\bf U} \partial_\mu - m \right ]
({\bf U}^{-1} \psi (x)) = 0 \quad \rightarrow \quad
\left [\gamma^5 \gamma^\mu\partial_\mu + m \right ]
\Psi (x) = 0\quad.\nonumber
\end{eqnarray}
According to the modern literature nothing would be changed.
Let us still assume that Prof. Dirac were found the equation
(\ref{eq}), but not Eq. (\ref{eqd}). What mathematical
framework were we have now?

\medskip

{\it Lagrangian.}

The first difference from the Dirac theory
arises when we try to construct the Lagrangian.

\smallskip

\noindent
{\tt Theorem:} There does not exist the Lagrangian
in terms of independent field variables $\Psi$ and $\overline \Psi$,
that could lead to the Lagrange-Euler equations of the form
(\ref{eq}) and (\ref{eqa}).

\smallskip

\noindent
{\tt Proof:}\footnote{{\it Cf.} with a consideration of vector
Lagrangians in ref.~\cite{Fush}.}  Most general form of the Lagrangian
in field variables $\Psi$ and $\overline \Psi$, $\partial_\mu \Psi$
and $\partial_\mu \overline \Psi$, that leads to the Lagrange-Euler
equations of the first order, is
\begin{eqnarray}
\lefteqn{{\cal L} = a_1 \overline \Psi \gamma^\mu \partial_\mu \Psi
+ a_2 \overline \Psi \gamma^\mu \gamma^5 \partial_\mu \Psi
+ a_3 \,\partial_\mu \overline \Psi \gamma^\mu \Psi
+ a_4 \,\partial_\mu \overline \Psi \gamma^\mu \gamma^5 \Psi +}\nonumber\\
&+& a_5 \overline \Psi \Psi + a_6 \overline \Psi \gamma^5 \Psi
+ a_{7\,\mu}\, \overline \Psi \sigma^{\mu\nu} \partial_\nu \Psi
+ a_{8\,\mu}\, \partial_\nu \overline\Psi \sigma^{\mu\nu} \Psi\quad.
\end{eqnarray}
Therefore,
\begin{eqnarray}
{\partial {\cal L}\over \partial \Psi} &=&
a_3 \partial_\mu \overline \Psi \gamma^\mu + a_4 \partial_\mu \overline \Psi
\gamma^\mu \gamma^5 +a_5 \overline \Psi +a_6 \overline \Psi \gamma^5
+a_{8\,\mu} \partial_\nu \overline \Psi \sigma^{\mu\nu}\quad,\nonumber\\
{\partial {\cal L}\over \partial \overline\Psi} &=&
a_1 \gamma^\mu\partial_\mu \Psi + a_2 \gamma^\mu\gamma^5\partial_\mu \Psi
+ a_5 \Psi +a_6 \gamma^5 \Psi
+a_{7\,\mu} \sigma^{\mu\nu}\partial_\nu \Psi\quad,\nonumber
\end{eqnarray}
\begin{eqnarray}
{\partial {\cal L}\over \partial (\partial_\mu \Psi)} &=&
a_1 \overline \Psi \gamma^\mu + a_2 \overline \Psi \gamma^\mu \gamma^5 +
a_{7\,\nu} \overline \Psi \sigma^{\nu\mu}\quad,\nonumber\\
{\partial {\cal L}\over \partial (\partial_\mu \overline \Psi)} &=&
a_3 \gamma^\mu \Psi + a_4  \gamma^\mu \gamma^5 \Psi +
a_{8\,\nu} \sigma^{\nu\mu} \Psi\quad.\nonumber
\end{eqnarray}
{}From the corresponding Lagrange-Euler equation for $\Psi$
one has $a_4 - a_2=1$ in order to obtain (\ref{eq}); from the adjoint
equation, $a_2 - a_4 =1$. Simultaneous satisfaction of these
equations is impossible. Theorem was proven.

How should we manage? Following to the logic of
the papers~\cite{Markov,Beli,Sokolik,DVA,DVO}
I propose to introduce the Dirac ``double",
another field that satisfies the equation:
\begin{equation}
\left [\gamma^5 \gamma^\mu \partial_\mu - m \right ]\Psi_2 (x) = 0\quad,
\label{equ2}
\end{equation}
and its adjoint
\begin{equation}
\overline \Psi_2 (x) \left [ \gamma^5 \gamma^\mu \loarrow{\partial}_\mu -
m \right ] = 0\quad.\label{eq2a}
\end{equation}
By using the concept of the two Dirac doubles
one can define, {\it e.g.}, the following Lagrangian:
\begin{eqnarray}
{\cal L} &=& {1\over 2} \left [ \overline \Psi_2
\gamma^\mu \gamma^5 \partial_\mu
\Psi_1 + \partial_\mu \overline \Psi_1 \gamma^\mu \gamma^5
\Psi_2 - \overline \Psi_1 \gamma^\mu \gamma^5 \partial_\mu \Psi_2 -
\partial_\mu \overline \Psi_2 \gamma^\mu \gamma^5 \Psi_1 \right ] -
\nonumber\\
&-& m \left [ \overline \Psi_1 \Psi_2 + \overline \Psi_2 \Psi_1
\right ]\quad,\label{lag}
\end{eqnarray}
which is Hermitian. It leads to the required equations
(\ref{eq},\ref{eqa},\ref{equ2},\ref{eq2a}).
Of course, the Lagrangian is defined within an
accuracy of the overall arbitrary constant term ({\it cf.}
Eq. (16) of ref.~[3b]). In this paper we choose it equal to the unit.

\medskip

{\it Relativistic covariance.}

In order the equation (for $\Psi_1$ or $\Psi_2$)
to be covariant it is necessary that
under Lorentz transformation $x^{\prime\,\mu} = L^\mu_{\,\,\nu}
x^\nu$ and $\Psi^\prime (x^\prime) = {\bf S} (L) \Psi (x)$ the primed
wave function satisfies the same equation:
\begin{equation}
\left [\gamma^5 \gamma^\mu \partial_\mu^{\,\,\prime} + m\right ]
\Psi^\prime (x^\prime) = 0
\end{equation}
or
\begin{equation}
{\bf S}^{-1} (L) \left [\gamma^5 \gamma^\mu
\frac{\partial x^\nu}{\partial x^{\prime\,\mu}}
\frac{\partial}{\partial x^\nu} +m \right ] {\bf S} (L) \Psi (x) = 0\quad.
\end{equation}
Therefore, we have
\begin{equation}
{\bf S}^{-1} (L) \gamma^5 \gamma^\alpha {\bf S} (L) =
L^\alpha_{\,\,\nu} \gamma^5 \gamma^\nu\quad.
\end{equation}
Restricting ourselves by
an infinitesimal proper transformation which may be
written as
\begin{equation}
L^\mu_{\,\nu} = g^\mu_{\,\,\nu} +\omega^\mu_{\,\,\nu}\quad,
\end{equation}
where the infinitesimal matrix $\omega_{\mu\nu}$ is
antisymmetric, one can obtain the same result that for the
Dirac theory. The generators of the Lorentz transformations
are
\begin{equation}
N^{\mu\nu}_{\Psi\Psi} \,=\, - \,{i\over 4} \sigma^{\mu\nu}\quad,\quad
N^{\mu\nu}_{\overline\Psi\, \overline \Psi} \,=\, + \,{i\over 4}
\sigma^{\mu\nu}\quad.
\end{equation}
Let us still not forget about the
possibility of combining the Lorentz and chiral transformations,
ref.~\cite{DVO}. It is the case when
the equation (\ref{eq}) comes over to Eq. (\ref{equ2}),
and reverse.  The transformed set of equations leaves to be unchanged.  In
this case the matrix of the transformation would be ${\bf S} (L) =
\gamma^5 \exp (-i\sigma^{\mu\nu}\omega_{\mu\nu}/4)$ and the generators are
\begin{equation}
N^{\mu\nu}_{\Psi_1 \Psi_2} \,= \,-\,{i\over 4} \gamma^5 \sigma^{\mu\nu}
\quad,\quad
N^{\mu\nu}_{\overline\Psi_1 \overline\Psi_2} \,=\, +\,{i\over 4}
\gamma^5 \sigma^{\mu\nu}\quad.
\end{equation}

\medskip

{\it Dynamical invariants.}

By means of the standard procedure~\cite{Bogol,Itzyk}
from the Lagrangian (\ref{lag}) one can
obtain dynamical invariants (that are deduced as
a consequence of the uniformity,
the isotropy of space-time and of the invariance of the Lagrangian
with respect to gradient transformations of the first kind):

\begin{itemize}

\item
The current vector and the charge operator:
\begin{eqnarray}
\lefteqn{J^\mu = i \sum_i \left [ \overline \Psi_i
\frac{\partial {\cal L}}{\partial
(\partial_\mu \overline \Psi_i )} -
\frac{\partial {\cal L}}{\partial (\partial_\mu
\Psi_i)} \Psi_i \right ]=\nonumber}\\
&=& i \left [ \overline \Psi_1 \gamma^\mu \gamma^5 \Psi_2
- \overline \Psi_2 \gamma^\mu \gamma^5 \Psi_1\right ]\quad,
\end{eqnarray}
\begin{equation}
Q \,=\, \int d^3 x \,J^{\,0} (x)\quad.
\end{equation}

\item
The energy-momentum tensor and the 4-vector of energy-momentum:
\begin{eqnarray}
\lefteqn{\Theta^{\mu\nu} = \sum_i \partial^\nu \overline \Psi_i
\frac{\partial {\cal L}}{\partial (\partial_\mu \overline \Psi_i)} +
\frac{\partial {\cal L}}{\partial (\partial_\mu \Psi_i)}\partial^\nu \Psi_i
- g^{\mu\nu} {\cal L} =\nonumber}\\
&=& {1\over 2} \left [ \partial^\nu \overline \Psi_1 \gamma^\mu\gamma^5
\Psi_2 + \overline \Psi_2 \gamma^\mu \gamma^5 \partial^\nu \Psi_1
- \partial^\nu \overline \Psi_2 \gamma^\mu \gamma^5 \Psi_1 -
\overline\Psi_1 \gamma^\mu \gamma^5 \partial^\nu \Psi_2 \right ]
- {\cal L} g^{\mu\nu}\quad,
\end{eqnarray}
\begin{equation}
P_\mu \,=\,\int d^3 x \,\Theta^{\,0}_{\,\,\mu} (x)\quad.
\end{equation}

\item
The angular-momentum tensor and the spin operator:
\begin{eqnarray}
\lefteqn{J^{\alpha,\mu\nu} = x^\mu \Theta^{\alpha\nu} - x^\nu
\Theta^{\alpha\mu} + 2 \sum_{ij}
\left [\frac{\partial {\cal L}}{\partial (\partial_\alpha \Psi_i)}
N^{\mu\nu}_{ij} \Psi_j + \overline \Psi_i N^{\mu\nu}_{ij}
\frac{\partial {\cal L}}{\partial (\partial_\alpha
\overline \Psi_j)}\right ] =} \nonumber\\
&=& x^\mu \Theta^{\alpha\nu} - x^\nu \Theta^{\alpha\mu}
+ {i\over 4} \left \{\overline \Psi_1\left [\gamma^\alpha
\gamma^5 \sigma^{\mu\nu} +\sigma^{\mu\nu} \gamma^\alpha \gamma^5\right
]\Psi_2 -\overline \Psi_2 \left [\gamma^\alpha \gamma^5 \sigma^{\mu\nu}
+\sigma^{\mu\nu} \gamma^\alpha \gamma^5 \right ] \Psi_1+\right.\nonumber\\
&+&\left . \overline \Psi_1 \left [\gamma^\alpha \sigma^{\mu\nu}
+\sigma^{\mu\nu}
\gamma^\alpha\right ]\Psi_1 - \overline \Psi_2
\left [\gamma^\alpha \sigma^{\mu\nu}
+\sigma^{\mu\nu} \gamma^\alpha\right ] \Psi^2\right \}\quad,
\end{eqnarray}
\begin{equation}
J^{\mu\nu} \,=\, \int d^3 x \,J^{\,0,\mu\nu} (x)\quad,\quad
(\hat W\cdot n) = S^{\,12}\quad,\quad\mbox{if}\quad {\bf n}\,\,\vert\vert\,\,
{\bf p} \,\,\vert\vert \,\,OZ \quad
\end{equation}
\end{itemize}
($S^{12}$ denotes the spin part of the angular momentum operator;
$\hat W_\mu$ is the Pauli-Lyuban'sky operator).

By using the plane-wave expansion:
\begin{mathletters}
\begin{eqnarray}
\Psi_1 (x) \,&=&\, \sum_h \int \frac{d^3 {\bf p}}{(2\pi)^3} {1\over 2p_0}
\left [ \Upsilon_h^{(1)} (p^\mu) \,a_h (p^\mu) \,e^{-i\,p\cdot x} +
{\cal B}_h^{(1)} (p^\mu) \,b_h^\dagger (p^\mu) \,e^{i\,p\cdot x}
\right ]\quad,\\
\Psi_2 (x) \,&=&\, \sum_h \int \frac{d^3 {\bf p}}{(2\pi)^3} {1\over 2p_0}
\left [ \Upsilon_h^{(2)} (p^\mu) \,c_h (p^\mu) \,e^{-i\,p\cdot x} +
{\cal B}_h^{(2)} (p^\mu) \,d_h^\dagger (p^\mu) \,e^{i\,p\cdot x}\right ]
\end{eqnarray}
\end{mathletters}
in the Fock space we obtain the following results:
\begin{equation}
\hat Q =  {1\over 2}\sum_h \int \frac{d^3 {\bf p}}{(2\pi)^3}\,{1\over 2p_0}
\left [a_h^\dagger (p^\mu) c_h (p^\mu) + c_h^\dagger (p^\mu)
a_h (p^\mu) - b_h (p^\mu) d^\dagger_h (p^\mu) - d_h (p^\mu)
b_h^\dagger (p^\mu)\right ]\quad,\label{charge}
\end{equation}
\begin{equation}
\hat {\cal H} = \hat P_0 = {1\over 2}\sum_h \int \frac{d^3 {\bf
p}}{(2\pi)^3} \,{1\over 2p_0}\, p_0 \left [ a_h^\dagger (p^\mu) c_h (p^\mu)
+ c_h^\dagger (p^\mu) a_h (p^\mu) + b_h (p^\mu) d_h^\dagger (p^\mu)  +
d_h (p^\mu) b_h^\dagger (p^\mu)\right ] \quad, \label{ham}
\end{equation}
and
\begin{eqnarray}
\lefteqn{(\hat W\cdot n) = {1\over 4} \sum_{h\,h^\prime} \int
\frac{d^3 {\bf p}}{(2\pi)^3}
{1\over 2p_0} \xi_h (\bbox{\sigma}\cdot {\bf n})
\xi_{h^\prime}\times}\nonumber\\
&\times&\left [ a_h^\dagger (p^\mu) c_{h^\prime} (p^\mu)
+ c_h^\dagger (p^\mu)
a_{h^\prime} (p^\mu) - b_h (p^\mu) d_{h^\prime}^\dagger (p^\mu)
- d_h (p^\mu) b_{h^\prime}^\dagger (p^\mu)  +\right.\nonumber\\
&+&\left. i\,a_h^\dagger (p^\mu) a_{h^\prime}
(p^\mu) - i\, c_h^\dagger (p^\mu) c_{h^\prime} (p^\mu) + i\,
b_h (p^\mu) b_{h^\prime}^\dagger (p^\mu) - i\, d_h (p^\mu)
d_{h^\prime}^\dagger (p^\mu)
\right ]\quad.\label{pl}
\end{eqnarray}
Like the usual formulation~\cite[p.145]{Itzyk} we have
to subtract the vacuum contributions in these dynamical invariants.
Next, let me note an interesting feature of this formulation.
For the first sight we obtained the positive-definite
Hamiltonian. Does this fact signify that there are no reasons for
introduction of anticommutation relations between operators $a_h (p^\mu)$
and $c^\dagger_h (p^\mu)$ (as well as between $b_h (p^\mu)$
and $d^\dagger_h (p^\mu)$)\,?

\medskip

{\it Propagators.}

Using the known procedure of St\"uckelberg and Feynman
for the field $\Psi_1$ (or $\Psi_2$)
one can obtain a local propagator, but it is not the Green's
function of the corresponding equation:
\begin{eqnarray}
\lefteqn{\left [\gamma^5 \gamma^\mu \partial^{\,x_2}_\mu
+ m \right ] \int \frac{d^{\,3}{\bf p}}{(2\pi)^3}\,
{1\over 2p_0} \left [a\,\theta (t_2 - t_1)
\sum_h \Upsilon_h (p^\mu) \otimes\overline \Upsilon_h (p^\mu)
e^{-ip\cdot (x_2 - x_1)}+\right.}\nonumber\\
&+& \left. b \,\theta (t_1 -t_2) \sum_h
{\cal B}_h (p^\mu) \otimes\overline {\cal B}_h (p^\mu)
e^{ip \cdot (x_2 -x_1)}\right ]
= {a\over 2}\gamma^5 \delta^{(4)} (x_2 -x_1)\quad,
\end{eqnarray}
provided that $a = -b$.
The same result is obtained for the second equation (\ref{equ2})
if we use the 4-spinors satisfying the second equation
(\ref{equ2})\footnote{The spinorial basis for 4-spinors satisfying the
second equation is chosen as
\begin{mathletters}
\begin{eqnarray}
\Upsilon^{(2)}_+ (\overcirc{p}^\mu) \,&=&\, \sqrt{{m\over
2}}\pmatrix{-i\cr 0\cr 1\cr 0\cr}\quad,\quad
\Upsilon_-^{(2)}
(\overcirc{p}^\mu) = \sqrt{{m\over 2}} \pmatrix{0\cr -i\cr 0\cr
1\cr}\quad,\quad\\
{\cal B}_+^{(2)} (\overcirc{p}^\mu) \,&=&\,
\sqrt{{m\over 2}}\pmatrix{i\cr 0\cr 1\cr 0\cr}\quad,\quad
{\cal B}_-^{(2)} (\overcirc{p}^\mu) =
\sqrt{{m\over 2}}\pmatrix{0\cr i\cr 0\cr 1\cr}\quad.
\end{eqnarray}
\end{mathletters}}
\begin{mathletters}
\begin{eqnarray}
\Upsilon^{(2)}_h (p^\mu) \,&=&\, -\,i\gamma^5 \Upsilon_h^{(1)}
(p^\mu)\quad,\quad {\cal B}_h^{(2)} (p^\mu) \,=\, i\,
\gamma^5 {\cal B}_h^{(1)} (p^\mu)\quad.\label{sp21}\\
\overline\Upsilon^{(2)}_h (p^\mu) \,&=&\, -\,i \overline \Upsilon_h^{(1)}
(p^\mu)\gamma^5\quad,\quad
\overline{\cal B}_h^{(2)} (p^\mu) \,=\, i\,\overline {\cal B}_h^{(1)}
(p^\mu)\gamma^5\quad.\label{sp22}
\end{eqnarray}
\end{mathletters}
We have used above that
\begin{mathletters}
\begin{eqnarray}
\sum_h \Upsilon_h^{(1)}
(p^\mu) \otimes\overline \Upsilon_h^{(1)} (p^\mu)= \sum_h {\cal B}_h^{(2)}
(p^\mu) \otimes\overline {\cal B}_h^{(2)} (p^\mu)= {1\over 2} \left [ \hat
p +im\gamma^5\right ]\quad,\\ \sum_h \Upsilon_h^{(2)} (p^\mu)
\otimes\overline \Upsilon_h^{(2)} (p^\mu)= \sum_h {\cal B}_h^{(1)} (p^\mu)
\otimes\overline {\cal B}_h^{(1)} (p^\mu)= {1\over 2} \left [ \hat p
-im\gamma^5\right ]\quad.
\end{eqnarray}
\end{mathletters}
Nevertheless,
taking into account that
\begin{mathletters}
\begin{eqnarray}
\sum_h\Upsilon_h^{(1)} (p^\mu)\otimes\overline {\cal B}_h^{(1)} (p^\mu)
\,&=&\,
-\,\sum_h {\cal B}_h^{(2)} (p^\mu) \otimes\overline\Upsilon_h^{(2)} (p^\mu)
={1\over 2}\left [\gamma^5 \hat p - i\,m\right ]\quad,\\
\sum_h{\cal B}_h^{(1)} (p^\mu) \otimes\overline \Upsilon_h^{(1)} (p^\mu)
\,&=&\,
-\,\sum_h \Upsilon_h^{(2)} (p^\mu) \otimes\overline {\cal B}_h^{(2)}
(p^\mu) = {1\over 2} \left [\gamma^5 \hat p + i\,m\right ]\quad,
\end{eqnarray}
\end{mathletters}
we obtain the needed result:
\begin{eqnarray}
\lefteqn{\left [\gamma^5 \gamma^\mu \partial^{\,x_2}_\mu +m \right ]
\int \frac{d^3 {\bf p}}{(2\pi)^3} {1\over 2p_0}\left [ a\,\theta (t_2 -t_1)
\sum_h \Upsilon_h ^{(1)} (p^\mu) \otimes\overline {\cal B}_h^{(1)} (p^\mu)
e^{-ip \cdot (x_2 -x_1)}+\right.}\nonumber\\
&+&\left . b\,\theta (t_1 -t_2) \sum_h {\cal B}_h^{(1)} (p^\mu) \otimes
\overline\Upsilon_h^{(1)} (p^\mu) e^{i\,p \cdot (x_2 -x_1)} \right ]
= \delta^{(4)} (x_2 -x_1)\quad,\quad\mbox{if}\quad a=-b=-2\,\, ,
\end{eqnarray}
and
\begin{eqnarray}
\lefteqn{\left [\gamma^5 \gamma^\mu \partial^{\,x_2}_\mu - m \right ]
\int \frac{d^3 {\bf p}}{(2\pi)^3} {1\over p_0}\left [ a\,\theta (t_2 -t_1)
\sum_h \Upsilon_h ^{(2)} (p^\mu) \otimes\overline {\cal B}_h^{(2)} (p^\mu)
e^{-ip \cdot (x_2 -x_1)}+\right.}\nonumber\\
&+&\left. b\,\theta (t_1 -t_2) \sum_h
{\cal B}_h^{(2)} (p^\mu) \otimes\overline \Upsilon_h^{(2)}
(p^\mu) e^{i\,p \cdot (x_2 -x_1)} \right ] = \delta^{(4)} (x_2 -x_1)\quad,
\quad\mbox{if}\quad a=-b=2\quad.
\end{eqnarray}
Finally,
\begin{mathletters}
\begin{eqnarray}
S_F^{(1)} (x) &=& \int \frac{d^4 p}{(2\pi)^4}
e^{-ip\cdot x} \frac{i\hat p \gamma^5 - m}{p^2 -m^2 +i\epsilon}\quad,\\
S_F^{(2)} (x) &=& \int \frac{d^4 p}{(2\pi)^4}
e^{-ip\cdot x} \frac{i\hat p \gamma^5 + m}{p^2 -m^2 +i\epsilon}\quad.
\end{eqnarray}
\end{mathletters}
The role of propagators is similar to the usual Dirac
theory: to propagate the positive frequencies toward positive times and
the negative ones, backward in time~\cite{FS}.

\medskip

{\it Physical contents and concluding remarks.}

First of all, I would like to note that one has a variety of
possibilities of physical interpretation of the results obtained in
the previous sections. It depends on the connection
between creation (annihilation) operators of the two fields and
setting the (anti)commutation relations between them.

Next, one has a puzzled problem with the imaginary part of
the Pauli-Lyuban'sky operator. It is very natural
to regard the particular cases when these imaginary terms
are cancelled each other. The creation and annihilation operators
of two parts of the doublet $\Psi_1$ and $\Psi_2$ are then
connected by the one of the following manners:

\begin{itemize}

\item
$c_h (p^\mu) = + a_h (p^\mu)$, \, $d_h (p^\mu) = - b_h (p^\mu)$;
or
$c_h (p^\mu) = - a_h (p^\mu)$, \, $d_h (p^\mu) = + b_h (p^\mu)$.
{}From the definitions of the plane-wave expansion and
from the relations (\ref{sp21},\ref{sp22}),
between 4-spinors $\Upsilon^{(1,2)}$ and ${\cal B}^{(1,2)}$
satisfying the first (\ref{eq})
and the second equation (\ref{equ2})
one follows that
$\Psi_2 (x) =\pm i\gamma^5 \Psi_1 (x)$.
If we set anticommutation relations
\begin{mathletters}
\begin{eqnarray}
\left \{ a_h (p^\mu), a_{h^\prime}^\dagger (q^\mu) \right \}_+ \,&=&\,
(2\pi)^3 \, 2p_0
\,\delta^{(3)} ({\bf p} -{\bf q}) \,\delta_{h\,h^\prime}\quad,\quad\\
\left \{ b_h (p^\mu), b_{h^\prime}^\dagger (q^\mu) \right \}_+ \,&=&\,
(2\pi)^3 \, 2p_0
\,\delta^{(3)} ({\bf p} -{\bf q}) \,\delta_{h\,h^\prime}\quad,
\end{eqnarray}
\end{mathletters}
we can assure ourselves that the formalism  describes
the charged particles of the opposite sign; the antiparticle
could be considered in the accordance with the common-used
``hole" interpretation. We have the Dirac's electron and positron
(or an muon and an antimuon), indeed.

\item
$c_h (p^\mu) = a_h (p^\mu)$, \, $d_h (p^\mu) = b_h (p^\mu)$.
Regarding the dynamical invariants (\ref{charge}-\ref{pl})
we obtain that the charge operator could be interpreted as the
particle number operator (there are no particles with the opposite
sign of the charge). The energy is not a positive-definite quantity
(antiparticles contribute to the energy with the opposite sign).
A compatibility of these two conclusions is unclear. Perhaps, this case
could be relevant to description of neutral particles.
If we were able to set the commutation ({\it not} anticommutation)
relations for this case, the physical interpretation
could be completely different.

\item
$c_h (p^\mu) = \pm i\,a_h (p^\mu)$, \, $d_h (p^\mu) = \pm i\,b_h (p^\mu)$.
Both the charge operator, the Hamiltonian and the Pauli-Lyuban'sky
operator are equal to zero. The interpretation of this physical
state is completely unclear (there are no particles at all?).
Perhaps, one can find some reminiscences with ``physical excitations"
discussed in ref.~[8d] and~\cite{AHL,Avdeev}.

\end{itemize}

I would also like to draw reader's attention to the problem of
the choice of field variables to use for the variation. It is easy to see
that, {\it e.g.}, under the substitution $\Psi_2 \rightarrow i\Psi_2$
the Lagrangian (\ref{lag}) is changed and it leads to the
different expressions for dynamical invariants\footnote{Of course, the
another expressions for dynamical invariants are connected
with the first ones by certain transformations
of creation (annihilation) operators.}.
Next, if start from the usual Dirac Lagrangian,
but vary using another field variables ({\it e.g.},
$\psi$ and $\gamma^5 \psi$ (or $\psi$ and $(1/ \sqrt{2})\,
(1-i\gamma^5)\,\psi$)
one can obtain physical excitations of the very different nature.
Could these physical excitations, following from the Dirac Lagrangian,
be relevant to describing neutrino (or, even, intermediate
vector bosons)?  Could we observe transitions between the Dirac
charged states described by Eq. (\ref{eqd}) and  the states described
by equations (\ref{eq},\ref{equ2}) ?
If yes, how should we correct the Feynman diagram technique?
It is undoubtedly that we would obtain the very different
expressions for self-energies, vertex functions and
for other diagrams involving the fields $\Psi_1$ and $\Psi_2$.
Finally,
Ryder~\cite{Ryder} writes:\, ``When a particle is at rest,
one cannot define
its spin as either left- or right-handed, so $\phi_R (0) = \phi_L (0)$."
In the papers~\cite{DVA,DVAG} it was paid attention to the more general
forms of the Faustov-Ryder-Burgard relation. Here we have considered
another form of the relation between left- and right- spinors at rest
and, as a result, we have deduced an interesting
model  based on this principle.
What physical excitations could we obtain
if set the Faustov-Ryder-Burgard
relation, {\it e.g.}, in the another form: $\phi_R (\overcirc{p}^\mu)
\,=\, (\pm i\sqrt{3}/2 -1/2)\,\phi_L (\overcirc{p}^\mu)$
({\it cf.} with the formulas (50) in ref.~[10b]) ?
Or, even in more general form?\footnote{For the discussion of
the most general case of the Faustov-Ryder-Burgard relation
see ref.~\cite{DVO953}.} Let us also note that the Lagrangian
(\ref{lag}) is compatible with a gauge principle.
So, the connection of the presented formalism with
non-Abelian gauge theories is transparent.
Therefore, it could serve as a ground for recreation of
the ideas proposed in the old papers~\cite{Utiyama,Kibble}.

\acknowledgments
This work is based on recent papers of Prof.  D. V.  Ahluwalia  and
fruitful discussions with Prof. A. F. Pashkov several years ago.
I appreciate encouragements of Prof. Yu. F. Smirnov
and Prof. I. G. Kaplan.
I thank Prof. A. Turbiner for informing me about the possibility
of the opposite sign at the mass term in the Dirac equation.
The questions of Prof. M. Moreno at the IFUNAM seminar (2/09/94)
were very helpful. I am obliged to Prof. R. N. Faustov for sending
me his paper and numerous useful advises.

I am grateful to Zacatecas University for professorship.

\end{document}